# ROLE OF DATA MINING IN NIGERIAN TERTIARY EDUCATIONAL SECTOR

**[1]Dauda Abdu, [2]Almustapha Abdullahi Wakili, [3]Lawan Nasiru, [4]Buhari Ubale**

[1]*Student, M.Tech. Computer Science & Engr., Department of Computer Science and Engineering, Mewar University Chittorgarh, bangis.ningi@gmail.com.*

[2]*Student, MSc. Software Engineering, Department of Computer Application, Mewar University Chittorgarh, almustaphawakili@gmail.com.*

[3]*Phd Research Scholar, Department of Computer Science Sangam University Bilwarah, lawannasiruhassan935@gmail.com.*

[4] *Student, MSc. Software Engineering, Department of Computer Application, Mewar University Chittorgarh, bugarko13@gmail.com.*

**ABSTRACT**

*Over a decade there has been a rapid growth in Nigerian educational system particularly higher education. Various institutions have come up both from public and private sector offering many of courses both under and post graduate students. Therefore, rates of students enroll for higher educational institutions in Nigeria have also increased. Hence it is very important to understand the roles plays by data mining in analyzing the collected data of students and their academic progression. It is a concern for today's education system and this gap has to be identified and properly addressed to the learning community. Data Mining it helps in various ways to resolve issues face in predictions students and staff performances within Nigerian education system. This paper work we discuss the roles of Data Mining tools and techniques which can be used effectively in resolving issues in some functional unit of Nigerian tertiary institutions.*

## 1. INTRODUCTION

The development of any country depends on the educational background level of its citizens. Seems to be an important matter for sustaining and developing the people thus it is among the early social service introduce. Most developing countries particularly Nigeria, experiencing increase in enrollment figure in tertiary educational system yearly, due to the increased quest for tertiary education by secondary school leavers. Despite the very fact that significant progress has been achieved over the years, absorption capacity is still low. Thus demand rate couldn't be overcome only by government Universities and colleges hence large numbers of personal private universities and institutions are being continuously establishes and the trend will continue within the future. As of last year (2020) report, the total number of Nigeria's universities is about 170 comprising of 43 Federal, 48 States and 79 Privates, 66.1% of the students are in Federal Universities while 27% are in the 48 State owned Universities and a meagre 6.9% are in the 79 Private universities (https://www.nuc.edu.ng). Nigeria have put more effort to ensure increases in number of present Universities and Colleges. In order to fill up the gap of present number of illiterate citizens which was reported in the 2018 that 13.6% of male in the urban areas are illiterate and up to 40.5% in the rural areas and that of female in urban areas about 26% are illiterate and 64.6% in rural areas.

Advent of information and communications technology (ICT) has completely transformed the ways in which data and information are created, structured, stored and accessed. Growing demand toward higher education led to increases in number of educational institutions, thus huge quantity of data is generated by every educational institution within each year. Hence, there is need of applying data mining techniques in order to transform this raw data into meaningful information. Data mining is the process of extracting hidden, unknown and potentially useful information and patterns from databases, data warehouses or other such data repositories.

Various tools and techniques are available that can help to improve the educational system by digitalizing educational data. Thus to simplify the ways in which researcher capture available data and retrieve or explore up to date information in order to make a proper decision.

In today's era educational systems try to offer standardized learning methods, by building a model of the individual's goals, attitude and knowledge. Educational Data Mining (EDM) can be seen as an application of data mining upon generated information from educational institutions.

In the educational field, data mining techniques can generate useful information and patterns that can be used for both educators and learners. EDM Not only may assist educators but also to improve the instructional materials in the field and to establish a decision process that will innovate the learning and teaching environment, but it may also provide recommendations to learners to improve their learning and to create individual learning environments.

## 2. RELATED WORK

Kumari, Abdul & Priyanka (2014) have discussed the evaluation of teaches' performance by using data mining techniques at higher institution and examining the role of Data mining in Educational Field by using SDAR to identify possible grade value like excellent, good, average and poor or fail. They also make used of K-means clustering algorithms for finding the best cluster center for attribute such as attendance, sessional mark and assignment mark. They have also discussed Rule-Based-Method (RBM) for extracting set of rule that show relationship between attributes of the data set and class level.

Srivastava & Abhay have examine the issue of predicting students' performance and their profiling through applying of data mining tools and techniques.

Astha, Vivek, Rajwant, & D., 2018) made an effort for finding the on student performance prediction with the help of classification models. A feature space is constructed by considering characteristics of family expenditure, family income, personal information and family assets of students. The potential/dominant features selection is unavoidable as it provides us with a subset of features. By using SVM classification algorithms we found our analysis very effective for our proposed features of family expenditure and student personal information categories. It can be easily derived from the results we got that academic information, family details and personal information have very strong impact on the students' performance due to instinctive reasons provided in discussions.

Tsai et al. (2011) examined computer proficiency test using data mining technique (cluster analysis, decision tree) to assess students' computer literacy before admitting into higher education. They concluded that data mining techniques assist universities to identify a number of groups who need reinforcement training and promote their computer proficiency more efficiently.

The most commercial data mining tools which provide multiple data mining functions and multiple knowledge discovery techniques and the environment on which they run are listed below:

| Name of tool and developer | Source (Commercial) | Function/Features | Techniques/Tools | Environments |
|---|---|---|---|---|
| MSSQL Server 2005 (Microsoft) An | Commercial | Provides MD functions Both in Relational DB system and Data warehouse (DWH) system environment | Integrates the algorithms by third party vendors and application users. | Windows, Linux |
| SPSS Clementine (IBM) | Commercial | Provides an integrated data mining development for end users and developers. | Association Mining, Clustering, Classification, Prediction and visualization tools | Windows, Solaris, Linux |
| Enterprise Miner (SAS Institute) | Commercial | Provides variety of statistical analysis tools | Association Mining, Classification, Regression, Time series analysis, Statistical analysis, Clustering | Windows, Solaris, Linux Insightful Miner |
| Oracle Data mining (Oracle Corporation) | Commercial | Provides an embedded DWH Infrastructure for multidimensional data analysis | Association Mining, Classification, Prediction, Regression, Clustering sequence similarity search | Windows, Mac, Linux |

| | | | and analysis. | |
|---|---|---|---|---|

**Table 1: commercial data mining tools**

### 3. DATA MINIG TECHNIQUES

In data mining many methods, techniques and algorithms are being applied in educational data mining (EDM) to achieve some goals. The most important methods that are usually used are Clustering, Classification and Prediction.

#### 3.1. Classification:

Classification algorithm is among the technique of data mining which helps to distinguish data classes and place data into appropriate predefined category. It is a supervised learning technique which needs categorized training data to come up with set of rules that will be used in mapping test data into pre-arranged category. It has two (2) phase of process, in which the first phase, it is a learning method, the training data sets are analyzed by the classification algorithm. The second phase it is where the classification test data sets are used to find the accuracy of the classification rules. If the accuracy is acceptable the rules can be applied to the new incoming data record.

#### 3.2. Prediction:

Regression technique is the method that can be used for prediction. Prediction is the based relationship between a thing that is known and a thing need to be predicted that is if certain attributes like domain knowledge and communication level of a student is known than his/her placement possibility can be predicted using multiple regression. Hence it refers to calculated assumptions for certain events made based on available processed data. Here placement possibility is dependent variable generally denoted by y and domain knowledge and communication level are independent variable generally denoted by x.

#### 3.3. Clustering:

Clustering can be defined as a data mining technique that can be used for making a group of similar classes of data. By using clustering technique, we can discover overall distribution pattern and correlations among data attributes. Clustering are classified into two methods based on it is cluster structure which provide partitioning and hierarchical cluster. So in this context of role played by data mining in educational system clustering can be used to classify students with similar characteristics in same cluster or group while removing others showing dissimilarities into some other clusters thus it can partition the similar groups from dissimilar groups by continuously measuring the Euclidean distance from cluster mean of similar groups.

## 4. APPLICATIONS OF DATA MINING IN EDUCATION SECTOR

Data mining have vital roles to play in some functional areas of educational sector such as.

a. **Predicting Students' Admission in Higher Education:** As more and more institutes are being establishes from both government and private sectors. Thus every year these institutions offer admissions for new students whereby, the admission process results in recording of huge volume amounts of data in institution's database. However, in most cases, this collected data is not put in a form of improving its use and results in wastage of what would otherwise be seen as the most precious assets of these institutions. Applying of some data mining techniques on this data, higher educational institutions can explore valuable information and predictions for the betterment of the admission process (Delavari & Beikzadeh, 2008). Prior to offer admission in any course some various factors need to be considered like secondary education performance. Yadav et al (2012) noted that the important attribute in predicting student's enrolment is found to be Graduation Stream (GS). The study shows that the student past academic performance can be used to create a decision tree model using ID3 algorithm that is used for predicting student's enrollment in postgraduate course. The Study found that students with background of mathematics or computers performed better in mathematical and computer course than students with other backgrounds. Below is a table that depict overall Nigerian federal, state and private Universities enrollment data for the year 2018, as reported by National University Commission (NUC). In order to analyze the enrollment data capacity at different level of Nigerian Universities.

| Institution | Enrollment Data |
|---|---|
| Federal University | 1146091 |
| State University | 1111823 |
| Private University | 541004 |
|  |  |

Table 2: Categories of Nigerian University Enrollment Data (2018)

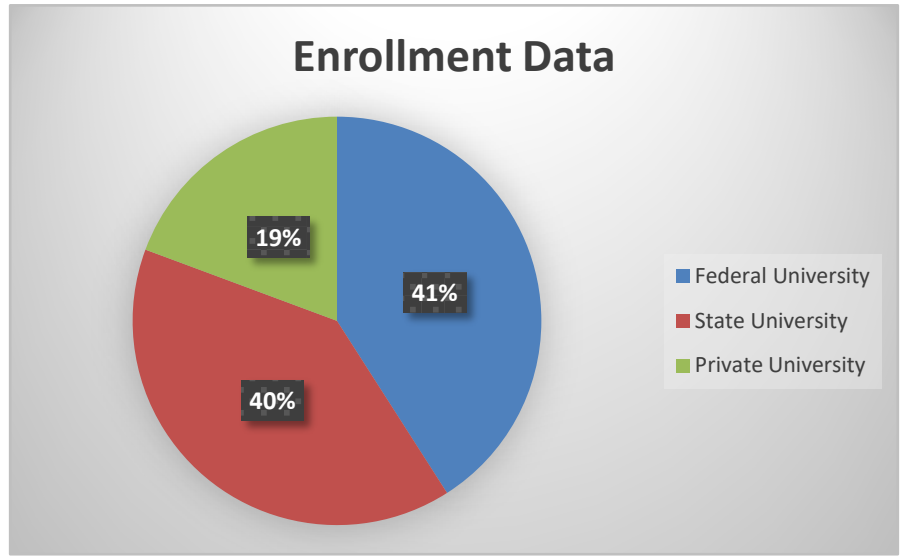

Figure 2: Categories of Nigerian University Enrollment Data 2018 (www.nuc.edu.ng)

**b. Predicting Students' Profiling:** Data mining technique can also be used as an effective tool in predicting profiling students based on both soft and hard skills. In this context the hard factors include academic background, grades and achievements while soft factor includes communication, behavior, attitude, hobbies etc. Naïve Bayes, Bayes Net, Support Vector Machines, Logistic regression and decision trees are the main data mining techniques and algorithm that are usually used for student profiling task. In addition, cluster analysis which is called data segmentation (Sinha et al., 2010), can also be used in predicting students' profiling.

**c. Predicting Students' Performance:** Data Mining is most popularly computing technique used to predict performance of students. Various prediction tools are available like regression and correlation analysis, decision tree, Bayesian networks, neural network etc. Mohammed & Alaa M. El-Halees (2012) have come up with a case study on how to apply different data mining techniques like Association Rule Mining, Clustering, Classification and Outlier Detection at various phases to improve the academic performance of graduate students.

Kumar and Uma (2009) studied students' performance in the course using data mining techniques, particularly classification techniques such as Naïve Bayes and Decision tree based on student ID and marks scored in course. In addition, many papers propose various techniques to improve the performance of students. Role of data mining techniques is not stick to predicting student performance only but can also be used in classifying teacher performance which helps in improving higher educational system.

Data mining processes it also helps students and teachers in order to improve students' performance and predict performance by applying two classification method Rule Induction and Naïve Bayesian classifier. To analyze the level of Nigerian graduate performance, data mining techniques was deploying, table below contain percentage of first class graduate for the year 2018 as reported by National Universities Commission. Below data was captured based on university level that is federal, state and private university.

| University | First Class Percentage |
|---|---|
| Federal University | 1.16 |
| State University | 0.93 |
| Private University | 5.7 |

Table 2: Nigerian Universities First Class Percentage(www.nuc.edu.ng)

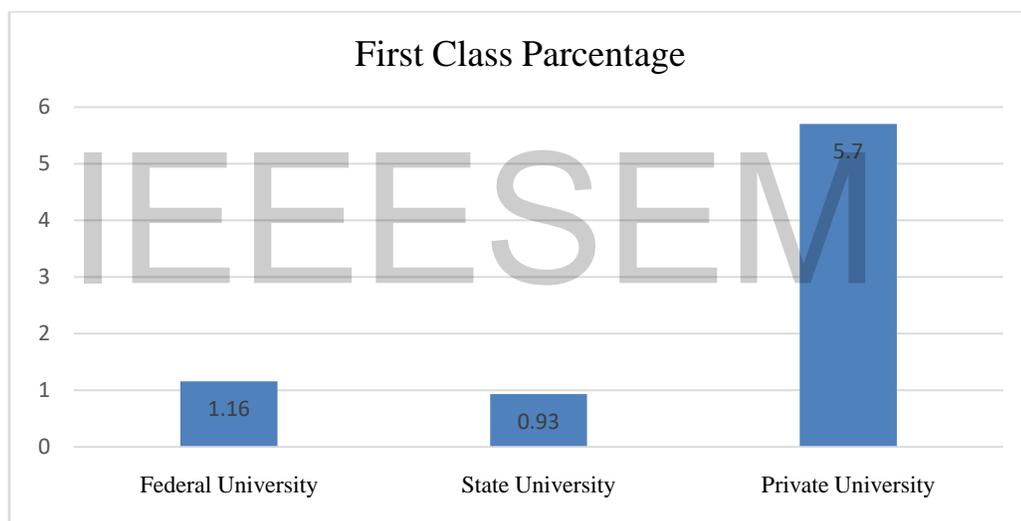

Figure 3: Percentage first class among 2018 graduate in Federal, State and Private Universities.

The above figure represents the percentage first class student in federal, state and private Universities of Nigeria.

d. **Teachers' teaching performance:** There can be various measures to judge teacher's teaching performance. Student feedback is a popular measure but often it gives skewed results. It is because there is high correlation found between marks of the student and feedback of the teacher. Several DM techniques have been used in this task, although association rule mining has been the most common. Mardikyan and Badur (2012) discover some factors which affect instructor performance in tertiary institution by using stepwise regression techniques of data mining. Instructor attitude, Teacher status, student attendance, and students' feedback affect teaching performance.

e. **Predicting Students' course selection:** Selection of course by a student depends on some various factors related to the student such as student's interest, course value grades etc. Kardan et al. (2013) discover some factors that affect the rate of student course selection using neural networks such as course type, course grades, course duration. The factors mentioned are used as an input of neural network modeling. Moreover, Guo (2010) analyze and make prediction about student course satisfaction by using neural networks. Reveal that number of students enrolled for a course and distinction rate in final grade are the must two influential factors that affect student satisfaction about a course.

## 5. CONCLUSION:

A lot of interest has been seen in EDM these days because a large number of students are enrolling for higher education. Through EDM institutions' researchers and stakeholders can bring more and more satisfaction amongst student's fraternity. Educational data mining finds its application not only in descriptive and predictive analytics but also in prescriptive analytics where suitable actions can also be prescribed. Understanding students, appropriate profiling and accurate predictions will not only increase the quality of education but also increase good learning experience to the students' fraternity.

Due to more and more usage of internet by students today, huge data is available about them. Through data mining, we can extract useful information that can help the education system to formulate appropriate strategies for our youths